\input harvmac
\input epsf

\def\ap{\alpha'}
\def\O{\Omega}
\def\p{\partial}
\def\da{^\dagger}
\def\oi{\oint}

\def\half{{1\over 2}}


\Title{}{\vbox{\centerline{The Fate of Massive F-Strings }}}

\centerline{Bin Chen,$^{1}$\footnote{}{Emails: bchen@itp.ac.cn,
mli@itp.ac.cn,  jhshe@itp.ac.cn}  Miao Li,$^{1,2}$  and Jian-Huang
She$^{2,3}$}

\medskip
\centerline{\it $^1$ Interdisciplinary Center of Theoretical Studies, Chinese Academy of Science}
 \centerline{\it P.O.Box 2735, Beijing 100080, P.R. China}

\medskip
\centerline{\it $^2$ Institute of Theoretical Physics, Chinese Academy of Science, }
 \centerline{\it P.O.Box 2735, Beijing 100080, P.R. China}
\medskip

\medskip
\centerline{\it $^3$ Graduate School of the Chinese Academy of Sciences, Beijing 100080, P.R. China}

\medskip

We calculate the semi-inclusive decay rate of an average string
state with compactification both in the bosonic string theory and
in the superstring theory. We also apply this calculation to a
brane-inflation model in a warped geometry and find that the decay
rate is greatly suppressed if final strings are all massive and enhanced
 with one final string massless, which may pose a challenge to this class of models.

\Date{April, 2005}

\nref\cosmic{M.B. Hindmarsh and T.W. Kibble, ``Cosmic strings",
Rept.Prog. Phys. 58, 477(1995); A. Vilenkin and E.P.S. Shellard,
{\it Cosmic strings and other topological defects}, Cambridge Unv.
Press (Cambridge 1994).}

\nref\coswit{E. Witten, ``Cosmic superstrings", Phys. Lett. B 153,
243(1985).}

 \nref\cosmica{E. J. Copeland, R. C. Myers and J. Polchinski,
``Cosmic F- and D-strings", hep-th/0312067; J. Polchinski,
``Introduction to Cosmic F- and D-strings", hep-th/0412244.}

\nref\cosmicb{G. Dvali, R. Kallosh and A. Van Proeyen, ``D-term
strings", JHEP 0401:035,2004, hep-th/0312005; G. Dvali and A.
Vilenkin, ``Formation and evolution of cosmic D-strings", JCAP
0403:010,2004, hep-th/0312007.}

\nref\sen{Ashoke Sen, ``Rolling Tachyon", JHEP 0204 (2002) 048,
hep-th/0203211.}

\nref\NSD{D. Kutasov, ``D-brane dynamics near NS5-branes",
hep-th/0405058.}

\nref\li{Bin Chen, Miao Li, Feng-Li Lin, ``Gravitational Radiation
of Rolling Tachyon",  JHEP 0211 (2002) 050, hep-th/0209222; Neil
Lambert, Hong Liu, Juan Maldacena, ``Closed strings from decaying
D-branes", hep-th/0303139;}

\nref\nsdecay{Y. Nakayama, Y. Sugawara and H. Takayanagi,
``Boundary states for the rolling D-branes in NS5 background",
JHEP 0407 (2004)020, hep-th/0406173;D. Sahakyan, ``Comments on
D-brane dynamics near NS5-branes", JHEP 0410
(2004)008,hep-th/0408070;Bin Chen, Miao Li and Bo Sun, ``Dbrane
Near NS5-branes: with Electromagnetic Field", JHEP 0412 (2004)037,
hep-th/0412022; Bin Chen and Bo Sun, ``Note on DBI dynamics of
Dbrane Near NS5-branes", hep-th/0401176; Y. Nakayama, K. L.
Panigrahi, S. J. Rey and H. Takayanagi, ``Rolling Down the Throat
in NS5-brane Background: The Case of Electrified D-Brane",
hep-th/0412038.}

\nref\openclose{A. Sen, ``Open-closed duality at tree level",
Phys. Rev. Lett. {\bf 91} (2003)181601, hep-th/0306137; Davide
Gaiotto, Nissan Itzhaki, Leonardo Rastelli, ``Closed Strings as
Imaginary D-branes", Nucl.Phys. B688 (2004) 70-100,
hep-th/0304192.}

\nref\kklt{Shamit Kachru, Renata Kallosh, Andrei Linde, Sandip P.
Trivedi, ``de Sitter Vacua in String Theory", Phys.Rev. D68 (2003)
046005, hep-th/0301240.}

\nref\kklmmt{Shamit Kachru, Renata Kallosh, Andrei Linde, Juan
Maldacena, Liam McAllister, Sandip P. Trivedi, ``Towards inflation
in string theory", JCAP 0310:013,(2003), hep-th/0308055 .}

\nref\rus{D. Amati and J. G. Russo, ``Fundamental Strings as Black
Bodies", Phys. Lett. B454(1999) 207 [hep-th/9901092].}

\nref\russ{Diego Chialva, Roberto Iengo, Jorge G. Russo, ``Search
for the most stable massive state in superstring theory", JHEP
0501 (2005) 001, hep-th/0410152 .}

\nref\ma{J. L. Manes, ``Emission Spectrum of Fundamental Strings:
An Algebraic Approach", Nucl. Phys. B621:37-61,2002
[hep-th/0109196].}

\nref\ingo{Roberto Iengo, Jorge G. Russo, ``The decay of massive
closed superstrings with maximum angular momentum", JHEP 0211
(2002) 045, hep-th/0210245;  Roberto Iengo, Jorge G. Russo,
``Semiclassical decay of strings with maximum angular momentum",
JHEP 0303 (2003) 030,hep-th/0301109;  Diego Chialva, Roberto
Iengo, Jorge G. Russo, ``Decay of long-lived massive closed
superstring states: Exact results", JHEP 0312 (2003) 014,
hep-th/0310283;  Diego Chialva, Roberto Iengo, ``Long Lived Large
Type II Strings: decay within compactification", JHEP 0407 (2004)
054, hep-th/0406271. }

\nref\life{D. Mitchell, N. Turok, R. Wilkinson and P. Jetzer,
``The Decay Of Highly Excited Open Strings", Nucl. Phys. B 315, 1
(1989) [Erratumibid. B 322, 628 (1989)];  J. Dai and J.
Polchinski, ``The Decay Of Macroscopic Fundamental Strings," Phys.
Lett. B 220, 387 (1989); H. Okada and A. Tsuchiya, ``The Decay
Rate Of The Massive Modes In Type I Superstring," Phys. Lett. B
232, 91 (1989); B. Sundborg, ``Selfenergies Of Massive Strings,"
Nucl. Phys. B 319, 415 (1989); D. Mitchell, B. Sundborg and N.
Turok, Nucl. Phys. B 335, 621 (1990).}

\nref\gsw{Michael B. Green, J.H. Schwarz, Edward Witten, SUPERSTRING THEORY. VOL. 1: Introduction; VOL. 2:
Loop Amplitudes, Anomalies and Phenomenology,
Cambridge, Uk: Univ. Pr. ( 1987) ( Cambridge Monographs On Mathematical Physics). }

\nref\joe{J. Polchinski, STRING THEORY. VOL. 1: An Introduction to the Bosinic string; VOL. 2: Superstring Theory
and beyond, Cambridge, UK: Univ. Pr. (1998) 531 p.}

\nref\ks{Igor R. Klebanov, Matthew J. Strassler, ``Supergravity
and a Confining Gauge Theory: Duality Cascades a nd
$\chi$SB-Resolution of Naked Singularities"JHEP 0008:052,(2000),
hep-th/0007191.}

\nref\recon{Mark G. Jackson, Nicholas T. Jones, Joseph Polchinski,
``Collisions of cosmic F- and D-strings". hep-th/0405229.}

\nref\hko{C. P. Herzog, I. R. Klebanov and P. Ouyang, ``D-branes
on the conifold and N=1 gauge/gravity dualities", hep-th/0205100.
}

\nref\gwvil{Tanmay Vachaspati, Alexander Vilenkin, "EVOLUTION OF COSMIC NETWORKS", Phys.Rev.D35:1131,1987.}

\newsec{Introduction}

The search for the imprints of the string theory in the cosmology
is one of the most important problems in string cosmology. Cosmic
string\cosmic, as a promising candidate, has drawn much attention
in the past few years. The renewed interest\cosmica\cosmicb in the
cosmic string resides in the fact that in the low-energy models
the cosmic strings could be created after inflation. This is in
contrast with the prediction of the old Planck-scale models, where
the cosmic strings produced before the inflation are ``blown
away"\coswit. It has been shown that some strings, including
fundamental string and solitonic strings, are sufficiently stable
and a few of them may survive and be observable in near future
through gravitational wave detectors such as LIGO and LISA.

In the string-inspired models on inflation, there is a class of
models based on  the brane-antibrane annihilation. In such
configurations, the tachyon field develop. As tachyon condense,
very massive closed string states are created, which could be an
origin of the cosmic string. However these massive fundamental
string states are not stable and may decay to other massive or
massless states. In order to understand the fate of a typical
massive F-string state better, it is essential to study its decay
systematically. On the other hand, even without considering its
phenomenological implication, the study of the decay of a typical
massive string state is an interesting issue on its own right in
string theory. From the investigation of the rolling tachyon \sen\
and the geometric tachyon realization in the (NS, D)-system\NSD,
it has been known that the tachyon condensation will radiate
various kinds of closed string states \li\nsdecay, even though
there does not exist a well-understood treatment of the
backreaction yet. To understand the tachyon condensation and the
mysterious open/close string duality\openclose, the study of the
fate of these massive string states is necessary. This is the
question we will address in this note.

The decay of massive strings is an old problem of string theory and
it is generally difficult to extract detailed knowledge from conventional calculations
mainly due to the exponentially growing state density. States at the same mass
level may be very different, especially their decay properties. For example,
there exist some peculiar string states, often with large angular momentum, which
are extraordinarily stable, with lifetime proportional to some positive power of their
masses \russ\ \ingo.

Fortunately in a realistic situation, such as the rolling tachyon
case, details of string states are not important, and  we are only
interested in some sort of averaged decay rates. A peculiar
long-lived string state may be created but with much smaller
probability, it must be in an extremal end of a distribution in
which the typical string states are amply produced and more
interesting. The lifetimes of the massive strings are cleverly
computed in this spirit \life. In \rus\ and \russ, massless string
emission from heavy strings were considered for both bosonic and
superstrings, where the authors average over the initial states of
the same mass level and fixed four momentum and sum over all
possible states for the remnant string, a thermal spectrum was
obtained. Manes considered in \ma\ the emission spectrum of the
strings of all mass levels for bosonic (open and closed) strings.
It was found there that the decay rates are universal, in the
sense that the leading terms depend only on the mass level of the
strings involved, and the more detailed properties of the states
affect only the sub-leading terms.

To study the stringy properties relevant for cosmology, we need to
consider the situation
 where some of the space-time dimensions are compactified. And in the realistic
 models
 building for early day inflation \kklmmt\ and present day accelerated expansion \kklt,
the superstrings rather than the bosonic strings are used. So in
this note, we consider first the decay process of the bosonic
strings with toroidal compactification in Section 2. Here the KK
modes and winding modes enter and make the process much more
complicated. And in Section 3, we go on to discuss the superstring
process in the flat space where the universal expressions similar
to \ma\ are found. We do compactification in Section 4 to
calculate the emission rate in a more realistic string inflation
model. Section 5 includes our conclusions and discussions.

\newsec{Bosonic string decay: with compactification}

In this section, we study the emission spectrum of massive closed bosonic strings in the case where
 $d_{c}$ out of the total $D$ dimensions are compactified on a flat torus.
A closed string state is characterized by its KK momentum numbers
$n_{i}$, winding numbers $w_{i}$, and its mass $M$ and momentum
$k$ in the noncompact dimensions. The mass shell condition reads
\eqn\mass{\ap M^2 = 4 (N_{R}-1)+ \ap Q_{+}^2 = 4 (N_{L}-1)+ \ap
Q_{-}^2} where $Q_{\pm} = \sum {n_{i}\over R_i}\pm {w_{i}R_i\over
\ap}$ , and $i=1,...,d_{c}$. For later convenience we also define
\eqn\mlr{4M_{RL}^2=M^2 -Q_{\pm}^2 .}

The decay amplitude for closed strings can be factorized into the
left and the right parts, and each part can be calculated in a way
similar to open strings. Consider first the  string
 of given mass $M$, winding $w_{i}$ and KK momentum $n_{i}$ in its center of mass frame decays into
two strings with mass $M_{1}$ and $M_{2}$, winding $w_{1i}$ and $w_{2i}$,
KK momentum $n_{1i}$ and $n_{2i}$, with similar mass shell conditions
\eqn\massl{\ap M_{1}^2 = 4 (N_{1R}-1)+ \ap Q_{1+}^2 = 4 (N_{1L}-1)+ \ap Q_{1-}^2,}
and
\eqn\massh{\ap M_{2}^2 = 4 (N_{2R}-1)+ \ap Q_{2+}^2 = 4 (N_{2L}-1)+ \ap Q_{2-}^2.}
In the following, we set $\ap = \half$. Energy conservation gives
\eqn\eng{M=\sqrt {M_1 ^2+k^2}+\sqrt{M_2^2+k^2},} with $k$ the
momentum in the noncompact dimension.

What we want to consider is the averaged semi-inclusive two-body
decay rate. That is, for the initial string, we average over all
states of some given mass, winding and KK momentum. For one of the
two final strings, we sum over all states with some given mass,
winding and KK momentum; only the other string's state is fully
specified (by keeping explicit its vertex operator).

This decay rate can be written as
\eqn\rate{\Gamma_{semi-incl} = {A_{D-d_{c}}\over {M^2}}g_{c}^2 {F_{L}\over{{\cal G}(N_L)}}
{F_{R}\over{{\cal G}(N_R)}}k^{D-3-d_c}\prod _i ^{d_{c}} R_{i}^{-1}}
with closed string coupling $g_c$, compactification radius $R_i$, and numerical coefficient
$A_p = {{2^{-p}\pi ^{{3-p}\over 2}}\over {\Gamma ({{p-1}\over 2})}}$, and $F_L$ and $F_R$ are given by
\eqn\ampl{F_{L}=\sum_{ \Phi_i|_{N_{L}}}\sum_{ \Phi_f|_{N_{2L}}} \big|
 \langle \Phi_f |  V_{L}(n_{1i},w_{1i},k)|\Phi_i \rangle \big|^2 }
\eqn\ampr{F_{R}=\sum_{ \Phi_i|_{N_{R}}}\sum_{ \Phi_f|_{N_{2R}}} \big|
 \langle \Phi_f |  V_{R}(n_{1i},w_{1i},k)|\Phi_i \rangle \big|^2 .}
The vertex operator for the one of the final state strings has
 been factorized into the left part $V_{L}$ and the right part $V_{R}$,
whose exponential parts look like $e^{ik_{L}\cdot X_{L}},
e^{ik_{R}\cdot X_{R}}$, with $-k_{L}^2 = M_{1}^2 -Q_{1-}^2$, which
we define to be $4m_L^2$ for later comparison with open strings.
In the flat spacetime the normalization of the initial state and
the final state two-body phase space contribute a factor
${k^{D-3}\over M^2}d\Omega$, while the integration over the solid
angle $\Omega$ gives $A_D$. Now momenta in the $d_c$ compact
dimensions are discrete, so we convert the integration to a
summation and this results in the factor $\prod _i ^{d_{c}}
R_{i}^{-1}$.

The sum in a fixed level is hard to perform, this can be done with
a trick: we insert a projector $P_{N}$ over states of level $N$,
and then sum over states of all levels, in the end we obtain a
trace \rus. Let \eqn\pro{P_N =\oint _C {dz\over z} z^ {\hat N-N}}
\eqn\proj{P_N \big| \Phi_{N'} \rangle = \delta _{NN'} \big| \Phi
_{N'} \rangle ,} so the amplitude squared can be written as
\eqn\trace{F_L = \oint _C {dz\over z}z^{-N_{L}} \oint _{C'} {dz'
\over z'}z'^{-N_{2L}} {\rm Tr} [z^{\hat N}V_{L}^\dagger(k,1)
z'^{\hat N}V_{L}(k,1)] ,} where we place the vertex operator at
$z=1$, takeing advantage of the $SL(2,R)$ invariance of the open
string tree amplitude. In the following, we define $w=zz'$, and
$v=z'$.

\subsec{The recursion relation}

Manes observed that \ma\ the trace in \trace\ describing a tree
level process can be carried out by converting it to the two point
correlation function on a cylinder with modulus $t$, where
\eqn\modulu{it=-{{2\pi i}\over{\ln w}}.} Note that \gsw\
\eqn\ttt{\int d^{D}p {\rm Tr} [z^{ L_{0} } V^\dagger (k,1){z'}^{
L_{0} } V(k,1)]=f(w)^{2-D}({{-2\pi}\over{\ln w}})^{D/2}< V^\dagger
(k,1) \ V(k,v) >, } where
\eqn\fw{f(w)=\prod_{n=1}^{\infty}(1-w^n)^{2-D}.} The second factor
comes from the  momentum integral over the zero modes, which does
not appear in the definition \ampl\ and \ampr\ of $F_{LR}$.
Remember that compactification does not affect the oscillation
modes, so we have \eqn\tra{{\rm Tr} [z^{\hat N}V_{L}^\dagger
(k_{L},1)z'^{\hat N}V(k_{L},1)] =f(w)^{2-D}
<{V'}_{L}^\dagger(k_{L},1) V'_{L}(k_{L},v)> ,} where the prime
means that $\partial _{\tau} ^{n}X^{\mu} (v)$ is replaced by
$v^{\hat N}\partial _{\tau} ^{n}X^{\mu} (1) v^{-\hat N}$ in the
vertex operator so that zero modes are excluded, and $D=26$ is the
full space-time dimension. Thus the $master formula$ for emission
of general states including KK modes and winding is
\eqn\master{F_L =\oint {dw\over w}w^{-N_{L}}f(w)^{2-D} \oint
{dv\over v}v^{N_L -N_{2L}} <{V'}_{L}^\dagger(k_L,1)V'(k_L,v)>.} It
can be  rewritten as \eqn\mast{F_L =\oint _{C_w}{dw\over
w}w^{-N_{L}}f(w)^{2-D} {{\cal I}_{N_{L}-N_{2L}}(w)}}
\eqn\rec{{\cal I}_n = \oint _{C_v}{dv\over v}v^n
<{V'}_{L}^\dagger(k_L,1)V'_L(k_L,v)>} where the contours satisfy
$|w| < |v| < 1$, for $|z|,|z'| < 1 $ . A general vertex operator
$V(k,z)$ is got from a physical state \eqn\phy{| \Phi > = \sum
c_{\mu _1 ... \mu _m}\alpha _{-n_1}^{\mu _1}...\alpha _{-n_m}^{\mu
_m}|0;k> } by replacing $\alpha _{-n}^{\mu}$ by ${i\partial
_{\tau}^{n}X^{\mu}(z)}\over{(n-1)!}$ , and $|0;k>$ by
$e^{ikX(z)}$, and the normal ordering is assumed. We take the
gauge\footnote{*}{Strictly speaking, such a choice of gauge only
exist in the critical dimension: for bosonic string $D=26$ and for
superstring $D=10$, which are implied in this
paper.}\eqn\gaug{k^{\mu _1}c_{\mu _1 \mu _2 ...\mu _m} = k^{\mu
_2}c_{\mu _1 \mu _2 ...\mu _m}= ... =k^{\mu _m}c_{\mu _1 \mu _2
...\mu _m} = 0 .} for which the contractions of exponentials with
$\partial X$ like terms vanish. Then the correlator in ${\cal I}_n
(w)$ factorizes into two parts, the contraction of two
exponentials and the contractions without exponentials.

Remember that $X$ scalar correlator on the cylinder is not changed
by compactification, so \eqn\opez{< e^{-ik_L X_L (1)} e^{ik_L X_L
(v)} >=e ^{-{1\over4}k_L ^2 \ln {\hat \psi (v,w)}} ={\hat \psi}
(v,w)^{m_L ^2},} \eqn\opep{< \partial
_{\tau}^{n}X_L^{\mu}(1)\partial _{\tau}^{m} X_L ^{\nu}(v) >=-(-)^n
\eta ^{\mu \nu} (v \partial _v)^{n+m}\ln \hat\psi (v,w), } with
\eqn\psih{\hat\psi(v,w)= (1-v)\prod_{n=1}^{\infty}
{(1-w^nv)(1-w^n/v)\over (1-w^n)^2} .} In the above correlators,
substitution has been made to account for
 the prime in the vertex operator. And $\partial _\tau$ is $v\partial_v$ for Euclidean proper
time $\tau = \ln z$. The correlator in ${\cal I}_n$ can thus be
written as \eqn\fac{< {V'}_L ^\dagger (k_L
,1)V'_L(k_L,v)>={\hat\psi}^{m_L ^2}{\cal P}(\Omega, \partial _\tau
\Omega  ...),} with \eqn\ome{\Omega (v,w)=<\partial
_{\tau}X_L(1)\partial _{\tau}X_L (v)
>=\sum_{n=1}^{\infty}n({v^n+{w^n(v^n+v^{-n})\over{1-w^n}}})}

The contour integration in ${\cal I}_n$ was computed in \ma. Since
the similar calculation is to be performed in the superstring
case, we in the following repeat the steps of \ma. Note that
\eqn\aaa{\hat\psi(w^{-1}v,w)=-{v\over w}\hat\psi(v,w) ,}
\eqn\bbb{\p_{\tau}^n\O(w^{-1}v,w)=\p_{\tau}^n\O(v,w) .} As
$m_L^2=2(N_L-1)$ is even, this makes the minus sign in \aaa\
irrelevant. The integration on the new contour $C'_v$ carries out
to be \eqn\ncon{\oint_{C'_v}{dv'\over v'}v'^n < {V'}_L\da
(k_L,1)V'_L(k_L,v') > =w^{-n-m_L^2}\oi _{C_v}{dv\over
v}v^{n+m_L^2}< {V'}_L\da(k_L,1)V'_L(k_L,v)  >, } which is just
\eqn\qqq{w^{-n-m_L^2} {\cal I}_{n+m_L^2}(w).} The difference
between the two contours is the integration over a small contour
around the singularity $v=1$ or $\tau=\ln v=0$. The function $\hat
\psi(v,w)$ has the behavior \ma\ \eqn\pvw{\hat
\psi(v,w)=-\tau(1+\half\tau)+O(\tau^3)}near $v=1$. So near $v=1$,
the correlators have the following asymptotic behavior
\eqn\exex{<e^{-ik_L X_L(1)}e^{-ik_L
X_L(v)}>=\hat\psi(v,w)^{m_L^2}=\tau^{m_L^2}(1+\half
m_L^2\tau+O(\tau^2)),} \eqn\parpar{<\p_{\tau}^nX_L(1)\p_{\tau}^m
X_L(v)>=-\p_{\tau}^{n+m-2}\Omega(v,w)=-{1\over{\tau^{n+m}}}+
O({1\over{\tau^{n+m-2}}}).} Note that one more partial derivative
for each vertex operator gives an additional $\tau^{-2}$ in
\parpar, but at the same time, the corresponding state is more
excited, contributing $\tau^2$ to \exex. The two factors exactly
cancel \eqn\coco{<V'^\dagger(k,1)V'(k,v)>={c\over{\tau^2}}(1+\half
m_L^2\tau)+O(1),} where the coefficient can be fixed to $c=1$ by
state normalization. So the contour integration around $v=1$ in
our case with compactification is the same as in the noncompact
case
 \eqn\scon{\oint_{C}{dv\over v}v^n < {V'}_L\da (k_L,1)V'_L(k_L,v) > = n+{m_L^2\over 2} .  }
We get the following recursion relation \eqn\recu{{\cal
I}_{n+m_L^2}(w) = w^{n+m_L^2}[{\cal I}_n (w)+n+{m_L^2\over 2}],}
which, used in sequence, can be recast into a series summation
\eqn\ssum{{\cal I}_n (w) = \sum_{p=1}^A
[n-m_L^2(p-\half)]w^{np-\half m_L^2(p^2-p)}+{\cal I}_\nu (w)w^{\nu
A+\half m_L^2(A^2+A)},} where $\nu = n-m_L^2A$, with $A$ an
integer. Then the contour integration in \master\ can be written
as \eqn\ppmm{F=\sum_{p=1}^A (n-m_L^2(p-\half)){\cal
G}[N_L-np+\half m_L^2(p^2-p)]+F_{NU},} where $n=N_L-N_{2L}$, and
\eqn\nuni{F_{NU}=\oi _{C_w}{dw\over w}w^{-N}f(w)^{2-D}{\cal I}_\nu
(w)w^{\nu A+\half m_L^2(A^2+A)},} and the generating function for
the mass level density \eqn\dens{{\rm Tr} w^{\hat
N}=f(w)^{2-D}=\sum_{N=0}^{\infty}{\cal G}(N)w^N .} has been used.

The Eq.\ppmm\ is useful in that all terms except the last one in
the decreasing series do not depend on the details of any of the
three string states involved (two are natural from the definition
eq.\ampl), thus knowing their levels is enough to determine them
fully. They are hence universal. The last term $F_{NU}$ is
non-universal. The recursion relation tells that the independent
functions are ${\cal I}_0, \cdots , {\cal I}_{m_L^2/2}$, which
depends on the particular vertex operators we inserted and so are
not universal. The contribution from $F_{NU}$ to $F$ is
generically negligible, in comparison with the universal part
contribution, except the case $A=0$ which happens when the emitted
states carry a large fraction of the total mass.

\subsec{The decay rate}

In the following, we shall calculate the decay rate \rate. In the
above discussions, we have fixed the levels of the incoming string
states and one of the outgoing string states, which are $N_{L,R},
N_{2L,2R}$ respectively. From the mass-shell conditions, we know
that once we fix the quanta of the incoming string states, the
outgoing string states could have various kinds of masses,
KK-momenta and windings, with respect to the energy condition
\eng, and conservations: \eqn\qq{Q_-=Q_{1-}+Q_{2-},~~~~
Q_+=Q_{1+}+Q_{2+}.} One important observation is that we have
inequality \eqn\ineq{\sqrt{N_{1L}}+\sqrt{N_{2L}} \leq
\sqrt{N_{L}}.} The equality saturate when \eqn\satu{k=0,\ \ \ \ \
\ {M \over Q_-}={M_2\over Q_{2-}},} where $k$ is the momentum in
the noncompact directions and $M_2, Q_{2-}$ are the quanta of the
outgoing string states with fixed level $N_{2L}$. The same
inequality holds in the right-mover.

Given a very massive initial string  of high level, its state
density has the asymptotic form \eqn\asy{{\cal G}(N)\sim \ N^{-
{D+1\over 4} } e^{a\sqrt{N} }\ ,\ \ \ \ \ a=2\pi\sqrt{D-2\over 6}\
.} The ratio between the first two terms in \ppmm\ can be
estimated to be \eqn\eee{{{\cal G}(N_L-n)\over{{\cal
G}(N_L-2n+m_L^2)} }\sim e^{a
\big({\sqrt{N_L-n}-\sqrt{N_L-2n+m_L^2}}\ \big)    }        .}

Using the inequality\ineq, it is at most of order
\eqn\ooo{\exp\big({a(\sqrt{N_L}-\sqrt{N_{2L}})}\big), \ \
\sqrt{N_{2L}}\geq{\sqrt{N_L}\over 2} \,} or
\eqn\oo{\exp({a(3\sqrt{N_{2L}}-\sqrt{N_{L}})}),\ \
\sqrt{N_{2L}}\leq {\sqrt{N_L}\over 2} .}
 In the extremal case $N_{2L}=N_L$, one can try to calculate ${\cal
I}$ directly. Thus if generically $N_L
> N_{2L}>{N_L\over 3}$, the first term dominates the whole summation in \ppmm, and the
other terms will be neglected to get \eqn\ppm{F_L \approx
(N_L-N_{2L}-\half m_L^2){\cal G}(N_{2L}).} $F_R$ can be carried
out in the same way.

Note that in our approximation, $F_{L,R}$ do not depend on the
details of the state specified in eq.\ampl\ and \ampr\ by the
vertex operators, and all states of the same level are emitted
with the same probability. Taking advantage of this, we can get
the total decay rate for decays into arbitrary states of given
mass, winding and KK momentum by simply multiplying eq.\rate\ by
the state density ${\cal G}(N_{1})$ \eqn\ttrate{\Gamma
[(M,n_i,w_i)\to(m,n_{1i},w_{1i})+(M_2,n_{2i},w_{2i})] \approx
A_{D-d_c}{g_c^2\over M^2}{\cal N}_L {\cal N}_R {\cal G}_L {\cal
G}_R k^{D-3-d_c}\prod _i ^{d_{c}} R_{i}^{-1},} where \eqn\nl{{\cal
N}_L=N_L-N_{2L}-\half m_L^2\ ,\ \ \ {\cal N}_R=N_R-N_{2R}-\half
m_R^2,} and \eqn\gl{{\cal G}_L={{{\cal G}(N_{1L}){\cal
G}(N_{2L})}\over{\cal G}(N_L)}\ ,\ \ \ {\cal G}_R={{{\cal G}
(N_{1R}){\cal G}(N_{2R})}\over{\cal G}(N_R)}.} Remember that
\eqn\mlr{m_{L}^2={1\over4}(m^2-Q_{1-}^2)\approx 2 N_{1L} . }


As long as $N_{1L}\gg 1$ and $N_{2L}\gg 1$, we can use eq.\asy\ to
write \eqn\gasy{{\cal G}_L \sim (2\pi T_H)^{-{{D-1}\over
2}}({N_{1L}N_{2L}\over {N_L}})^{-{{D+1}\over
4}}e^{-\sqrt{2}t_L/T_H},} with the Hagedorn temperature
$T_H={1\over\pi}\sqrt{3\over{D-2}}$ and
$t_L=\sqrt{N_{L}}-\sqrt{N_{1L}}-\sqrt{N_{2L}}$ in a sense coming
from the kinetic energy released in the decay process. We have in
the above restored the multiplicative constant in front of the
state density. Note that for later convenience, in our notation we
set ${\cal G}(M)dN={\cal G}(N)dN$, a little different from usual
sense ${\cal G}(M)dM={\cal G}(N)dN$.

>From the discussion on the inequality \ineq, we know that the
dominant decay channel is the original incoming string break into
two strings with the same \eqn\ratio{\alpha={M\over Q_-}={m\over
Q_{1-}}={M_2\over Q_{2-}}} and without releasing any kinetic
energy $k=0$. In other words, the outgoing string states incline
to carry proportional KK-momenta and windings to their masses.
Furthermore, in such string breaking without releasing the kinetic
energy, it is least probable that the original string break into
two strings with the same mass and quanta.

 To see more clearly the
decay process, we may take into account the dependence on the
momentum $k$ and other quanta. An efficient way is to expand
around the maximum of the exponential. It turns out to be
\eqn\k{{\cal G}_L \sim \exp({-{\sqrt{2}k^2\over
16T_H}\sqrt{{N_L\over N_{2L}}}{1\over
\sqrt{N_L}-\sqrt{N_{2L}}}})\exp({-\sqrt{2N_LN_{2L}}\epsilon^2\over
(\alpha^2-1)^2(\sqrt{N_L}-\sqrt{N_{2L}})2T_H})} where $\epsilon$
characterize the deviation of ${M_2\over Q_{2-}}$ from $\alpha$.
Thus the dependence on $k$ is still a Gaussian distribution, which
could be reduced to the one in \ma  without compactification. The
influence of the KK-momenta and windings is the change of the
coefficients in the exponential
\eqn\change{\exp({-{\sqrt{2}k^2\over 16T_H}\sqrt{{N_L\over
N_{2L}}}{1\over \sqrt{N_L}-\sqrt{N_{2L}}}})=\exp({-{\alpha \over
\sqrt{\alpha^2-1}}{k^2\over 2T_H}{M\over 2mM_2}})} which tell us
when $\alpha$ is huge, namely the original windings and
KK-momentum is small, the distribution reduces to the one without
compactification. Moreover, the contribution of KK-momentum and
windings has also a Gaussian distribution, and the total
contribution could roughly be integrated out and give us a finite
number. However, things are more interesting here: from the
distribution of the $\epsilon$, it seems that when $\alpha
\rightarrow \infty$, which happens at $Q_-=0$, all kinds of quanta
of $Q_{2-}$ could give the same contribution. This is just an
illusion, due to the subtlety on the expansion. In this case, the
outgoing strings tends to carry very small $Q_{2-}$ quanta, with a
distribution as $\exp(-{MM_2 \over m}{\delta^2 \over 2T_H})$,
where $\delta$ characterize the deviation from the extreme. In any
case, one may take into account of all the contributions of the
KK-momenta and windings well approximately by integrating over the
distributions above.


With the above assumptions, eq. \nl\ is approximately
\eqn\nlapp{N_L-N_{2L}-\half m_L^2\approx {\alpha^2-1\over
\alpha^2}{mM_2 \over 4}, \quad N_R-N_{2R}-\half m_R^2\approx
{\beta^2-1\over \beta^2}{mM_2 \over 4},} where $\beta={M \over
Q_+}$.

Now gathering all above together, we can calculate the total rate for a string with mass $M$, KK momentum $n_i$,
 winding $w_i$ to produce a string with mass $m$,
 by integrating over noncompact momentum $k$ (or say it in another way, summing over the other final string's mass level),
 and summing over partitions $(n_{1i},n_{2i})$ and $(w_{1i},w_{2i})$ with
constraints $n_i=n_{1i}+n_{2i}$ and $w_i=w_{1i}+w_{2i}$.

First, we do the summations
\eqn\emisrt{\Gamma_m(M,n_i,w_i)=\sum_{N_2}\sum_{(n_{1i},n_{2i})}\sum_{(w_{1i},w_{2i})}\Gamma
[(M,n_i,w_i)\to (m,n_{1i},w_{1i})+(M_2,n_{2i},w_{2i})].} To
illustrate the picture more clearly, let us consider a special
case, where the initial string carry very small $Q_-$. In this
case, ${\cal G}_L{\cal G}_R$ can be divided into three parts: the
kinetic energy, the KK modes and winding modes \eqn\gleft{{\cal
G}_L{\cal G}_R \sim {\cal G}^k {\cal G}^{KK} {\cal G}^w , } where
\eqn\glk{{\cal G} ^k=\exp[{-{k^2\over {2T_H}} {M\over{m M_2}} }]
,} \eqn\glkk{{\cal G} ^{KK}= \exp[{-{1\over{2
T_H}}\sum_{i=1}^{d_c} {1\over{R_i^2}}({n_{1i}^2\over
m}+{n_{2i}^2\over M_2}-{n_i^2\over M} )  }], } \eqn\glw{{\cal G}
^w= \exp[{-{2\over{T_H }}\sum_{i=1}^{d_c}R_i^2 ({w_{1i}^2\over
m}+{w_{2i}^2\over M_2}-{w_i^2\over M} )  }].} We see from eq.\glk\
that the probability to emit large kinetic energy strings is
exponentially suppressed as expected. The factor \glkk\ from the
KK modes, when we fix $n_i=n_{1i}+n_{2i}$, has a saddle point
which means that the processes satisfying ${n_{1i}\over
m}={n_{2i}\over M_2}$ are much preferred. Similarly the saddle
point for the winding modes occurs at ${w_{1i}\over
m}={w_{2i}\over M_2}$. That is, for the final two strings, the
heavier one tends to carry larger KK momentum and more windings.
This is consistent with our general argument above.  The decay
rate can correspondingly be factorized into three parts
\eqn\emrtf{\Gamma_m(M,n_i,w_i)=\Gamma^k
\sum_{(n_{1i},n_{2i})}{\cal G}^{KK}\sum_{(w_{1i},w_{2i})}{\cal
G}^w,} where we make saddle point approximation for the noncompact
momentum, that is we let $k=0$, or $M_2=M-m$,
 when considering the KK modes and winding modes. Note that the
basic $\vartheta$ function has the
expansion\eqn\thefun{\vartheta(\nu|\tau)\sum_{n=-\infty}^{\infty}\exp(\pi in^2\tau+2\pi in\nu),} so the
two summations in \emrtf\ which are actually summations over
single argument $\sum_{n_{1i}=-\infty}^{\infty}$ and
$\sum_{w_{1i}=-\infty}^{\infty}$, can be written more concretely
in terms of $\vartheta$ functions
\eqn\nsumthe{\sum_{(n_{1i},n_{2i})}{\cal G}^{KK}\approx
\prod_{i=1}^{d_c} e^{\beta_i}\vartheta(\nu_i|\tau_i), \quad
\sum_{(w_{1i},w_{2i})}{\cal G}^w\approx
\prod_{i=1}^{d_c}e^{\beta'_i}\vartheta(\nu'_i|\tau'_i),} where
\eqn\betao{\beta_i=-{1\over{2T_H R_i^2}}{m\over{M(M-m)}}n_i^2,}
\eqn\betat{\beta'_i=-{{2R_i^2}\over{T_H
}}{m\over{M(M-m)}}\omega_i^2 ,} \eqn\nuone{\nu_i={1\over{2\pi i
T_H R_i^2}}{n_i\over{M-m}},} \eqn\nutwo{\nu'_i={{2R_i^2}\over{\pi
i T_H }}{\omega_i\over{M-m}}   ,} \eqn\tauon{\tau_i={i\over{2\pi
T_H R_i^2}}{M\over{m(M-m)}},}
\eqn\tautw{\tau'_i={{2iR_i^2}\over{\pi T_H }}{M\over{m(M-m)}} .}

In general, as we argued above, the contribution from the KK-modes
and windings could be just a constant and the dependence on the
momentum $k$ could be encoded in \eqn\mom{{\alpha^2-1\over
\alpha^2}{\beta^2-1\over \beta^2}({mM_2 \over
4})^2\exp\big({-({\alpha \over \sqrt{\alpha^2-1}}+{\beta \over
\sqrt{\beta^2-1}}){k^2\over 2T_H}{M\over 2mM_2}}\big).}Note that
if we measure the energy of the one of the decay string with mass
$m$, we find a temperature: \eqn\tem{T_m={T_H\over 2}({\alpha
\over \sqrt{\alpha^2-1}}+{\beta \over \sqrt{\beta^2-1}})^{-1}(1-{m
\over M}).} This reflects both the recoil effect and the influence
of the KK-modes and windings.

It is remarkable that when $\alpha=1$ or $\beta=1$, the decay is
completely suppressed. This interesting limit corresponds to the
BPS condition $M=Q_-$ or $M=Q_+$ in the superstring case. As we
show in the next section, all the discussion here could be applied
to the superstring. Therefore, the fact that the BPS string states
are stable is reflect in the above special limit. Actually, the
suppression not only happens in the decay to the final massive
strings, it also happens in the thermal radiation of the massless
particles\rus. Moreover, from \mom, the decay of the near BPS
string states are greatly suppressed.

 The measure of the integral in $\Gamma^k$ is $dN_2\sim M_2
dt$,
 where we define
 \eqn\t{t=({\alpha \over \sqrt{\alpha^2-1}}+{\beta \over
\sqrt{\beta^2-1}}){k^2\over 2T_H}{M\over 2mM_2}.}Expressed in an
integral over $t$ \eqn\gggam{\Gamma^k\approx const. A'_{d_c} g_c^2
{(2\pi T_H)}^{-D+1}{M\over m}
m_R^{-{{D-1+d_c}\over2}}\int_0^{\infty}
t^{{D-3-d_c}\over2}e^{-{t\over T_H}}dt,} where $m_R={{m(M-m)}\over
M}$, and $A'_{d_c}=\prod_{i=1}^{d_c} R_i^{-1}$. Thus including the
phase space factor, we have a Maxwell-Boltzmann distribution for
$t$, which could be taken as a measure of the total kinetic
energy. Also the mean kinetic energy released per decay could be
characterized by \eqn\kin{<t>={D-1-d_c \over 2}T_H,} which is
independent of the masses and satisfies the equipartition
principle in $D-1-d_c$ noncompact spatial dimensions.
 Carrying out the integral, which is just a $\Gamma$-function, we get
\eqn\gggamm{\Gamma ^k\approx A'_D g_c^2{M\over
m}m_R^{-{{D-1+d_c}\over2}},} with $A'_D= const.
A'_{d_c}{T_H}^{-{(D-1+d_c)\over2}}$.

Then, we take into account the fact that there are $\rho(m)dm$ levels between mass $m$ and $m+dm$, with $\rho(m)={1\over4}m$,
so the total emission rate of strings of mass $m$ is
\eqn\diffrt{{d\Gamma(m)\over{dm}}=\Gamma_m(M,n_i,w_i)\rho(m)={{d\Gamma^k(m)}\over{dm}} \sum_{(n_{1i},n_{2i})}{\cal G}^{KK}\sum_{(w_{1i},w_{2i})}{\cal G}^w,}
where \eqn\diffmn{{{d\Gamma^k(m)}\over{dm}}=\Gamma^k \rho(m)=A''_D g_c^2Mm_R^{-{{D-1+d_c}\over2}},}
with $A''_D={1\over4}A'_D$. It can easily be checked that if we set $d_c=0$,
 the above result reduces to the flat space case considered in \ma\ as
 expected. However, here is a puzzle. Naively, from semiclassical point of view, one may
 expect to
 interpret the probability for two points with distance $m$ on the string to
 meet is inversely proportional to the volume $m^{D-1-d_c/2}$, where $m^{1/2}$ is the mean distance between two points
 and $D-1-d_c$ is the noncompact spatial dimensions. In other
 words, the less the noncompact spatial dimensions, the more
 possible the two points on the string to meet, inducing the
 breaking of the string. But \diffmn\ tells that this is not true:
it seems that  the more we compact the dimensions, the less
possible is the string break.

\newsec{Superstring decay: flat space}

In this section, we calculate the decay rate of a typical massive type II
string state in the 10-d dimensional Minkowski space. We consider
first the emission of NS-NS states, which we will argue later is the dominant decay channel.
The basic quantity
 to compute is one piece of the semi-inclusive amplitudes squared as defined in \ampl\ and \ampr,
\eqn\amps{F={1\over {\cal G} (N)}\sum_{ \Phi_i|_N}\sum_{ \Phi_f|_{N'}} \big|
 \langle \Phi_f |  V(k)|\Phi_i \rangle \big|^2. }
Carry out the projection
\eqn\amptr{F={1\over\cal{G}(N)}\oint_{C}{dz'\over
z'}z'^{-N'}\oint_{C'}{dz\over z}
 z^{-N}{\rm Tr}[{{1 +e^{i \pi F}}\over 2}V^\dagger (k,1){{1 +e^{i \pi F}}\over 2}z'^{\hat N}V(k,1)z^{\hat N}],}
 where $F$ is the world sheet fermion number.

We will mostly follow the procedure of \ma, and the main task is to
 check whether there is still a recursion relation when the fermions are
 included.

Convert the above tree level trace to a cylinder correlator
\eqn\strace{{\rm Tr}[{{1 +e^{i \pi F}}\over 2}V^\dagger (k,1){{1
+e^{i \pi F}}\over 2}z'^{\hat N}V(k,1)z^{\hat N}]=f_{NS}(w)
<V'^\dagger(k,1){{1+e^{i\pi F}}\over 2}V'(k,v)>,} where
\eqn\nsf{f_{NS}(w)=\sum _{n=0}^{\infty}{\cal G}_{NS}w^n={1\over
\sqrt w}{\rm Tr}[{{1+e^{i \pi F}}\over 2} w^N],} with ${\cal
G}_{NS}$ the state density of the NS sector. Note that now
 the massless level already involves $1\over 2$
unit of excitation, so the factor $1\over \sqrt w$ should be
included.

We define ${\cal I}_n$ in the same way as in \rec. On the cylinder
we use vertex operators in the $0$ picture in the notation of
\joe. For example, the massless state is \eqn\vortz{V_0 =(\xi .\p
X +i \xi .\psi k .\psi)e^{ik.X}.} The state generated by $\psi
_{-{3\over 2}}^\mu $ has the vertex operator \eqn\votth{V_1 =(\xi
.\p ^2 X+i\xi .\p \psi k.\psi)e^{ik.X}.} More highly excited
states can be constructed with higher derivatives and more terms
with building blocks as above.

These facts suggest that the correlator in \strace\ can be factorized as in \fac,
imposing the same gauge conditions. Now the polynomial is also a
function of the fermion correlators and their derivatives
\eqn\sfac{< {V'} ^\dagger (k ,1)V'(k,v)>={\hat\psi}^{m ^2}{\cal
P}(\Omega,
\partial _\tau \Omega  ...;\Lambda,\partial _\tau \Lambda,...),}
where we define \eqn\fcorr{\Lambda (v,w) =<\psi(1)
\psi(v)>.}Actually there are two subtleties in arriving \sfac. One
is that though the vertex operator of the massive string state
might be just the linear combinations of the terms which are the
product of chiral and anti-chiral parts, the contribution from the
contraction of two exponentials is universal and the end result
could still be written as \sfac. The other subtlety is the gauge
choice which allow the contraction between the exponential and the
terms without exponential vanishing. As in bosonic case \gaug,
such a gauge is believed to exist in the critical dimension.

What is new for the superstring case is that due to the presence
of fermions, the correlator in \strace\ is a summation over spin structures.
There are four spin structures on the cylinder. Restoring spin structures,
 the above correlator or Green function  can be written as
\eqn\green{G_{(a,b)}(w_1-w_2)=<\psi(w_1)\psi(w_2)>_{(a,b)},} with
$a,b=0,\half$ labelling the periodicity
\eqn\grpre{G_{(a,b)}(z+2\pi)=e^{2i\pi a}G_{(a,b)}(z), \quad
G_{(a,b)}(z+it)=e^{2i\pi b}G_{(a,b)}(z).}

The Green functions for the so-called even spin structures can be explicitly written as
 \eqn\fone{<\psi ^\mu (w_1)\psi ^\nu
(w_2)>_{(0,\half)}={\ap\over{4\pi}}g^{\mu\nu}{{\vartheta_2\bigl({{w_1-w_2}\over{2\pi}}|it\bigr)\vartheta_1'(0|it)}\over
        {\vartheta_1\bigl({{w_1-w_2}\over{2\pi}}|it\bigr)\vartheta_2(0|it)}},}
\eqn\ftwo{<\psi ^\mu (w_1)\psi ^\nu
(w_2)>_{(\half,\half)}={\ap\over{4\pi}}g^{\mu\nu}{{\vartheta_3\bigl({{w_1-w_2}\over{2\pi}}|it\bigr)\vartheta_1'(0|it)}\over
        {\vartheta_1\bigl({{w_1-w_2}\over{2\pi}}|it\bigr)\vartheta_3(0|it)}},}
\eqn\fthree{<\psi ^\mu (w_1)\psi ^\nu
(w_2)>_{(\half.0)}={\ap\over{4\pi}}g^{\mu\nu}{{\vartheta_4\bigl({{w_1-w_2}\over{2\pi}}|it\bigr)\vartheta_1'(0|it)}\over
        {\vartheta_1\bigl({{w_1-w_2}\over{2\pi}}|it\bigr)\vartheta_4(0|it)}},}
We can explicitly check their periodicity by using properties of $\vartheta$ functions
\eqn\periodo{\vartheta _1(\nu+1|\tau)=-\vartheta_(\nu|\tau),}
\eqn\periodt{\vartheta _2(\nu+1|\tau)=-\vartheta_2(\nu|\tau),}
\eqn\periodth{\vartheta _3(\nu+1|\tau)=\vartheta_3(\nu|\tau),}
\eqn\periodf{\vartheta_4(\nu+1|\tau)=\vartheta_4(\nu|\tau).}
These three spin structures are permuted under modular transformations,
so to ensure modular invariance they should always be kept at the same time.

For the doubly periodic case $(a=0,b=0)$, which is often called
odd spin structure, the existence of zero modes of the Dirac
operator make things subtle(see for example \joe). It is modular
invariant itself and does not mix with the other three under
 modular transformation. And it is well known that this sector does not
contribute to the correlation functions if the insertion of the
vertex operator cannot eat the zero modes, while in our
consideration of very massive strings, the corresponding vertex
operators can swallow the zero modes without trouble.

Note that in our convention, the arguments in the $\vartheta
(\nu|\tau)$ function read \eqn\aug{\nu={{\ln v}\over{\ln w}},\quad
 \tau=-{{2\pi i}\over{\ln w}} .} To get a second contour we make the
 transformation $v'=w^{-1}v$, which amounts to replace $\vartheta (\nu|\tau)$
 by $\vartheta (\nu-1|\tau)$, or $G_{(a,b)}(z)$ by $G_{(a,b)}(z-2\pi)$.

Using \grpre\ we get
\eqn\lamd{\Lambda(w^{-1}v,w)=\pm\Lambda(v,w).}
 The minus sign in \lamd\ is irrelevant since we always have even numbers of
fermions in our vertex operators after GSO projection.

Taking derivatives of $\vartheta$ functions we can see that
 derivatives of $\Lambda$ get at most an extra
irrelevant minus sign under the transformation $v'=w^{-1}v,w'=w$.
 So it turns out that the
 fermion correlators  have the same good property as
 $\p X$ correlators that they (effectively) do not change under our transformation. This is
 after all not a surprise since we know, for example from bosonization, that two fermions
  correspond to the derivative of a bosonic field, and GSO projection
enables us to always have paired world-sheet fermions.

The contour integrations on contour $C'_v$ and around the
singularity $v=1$ both carry on to the superstring case. The
latter integration depends only on the asymptotic behavior of the
correlators of $e^{ikX}$, $\p X^{\mu}$ and $\psi\psi$ near $v=1$,
while the last two have the same order dependence on $\tau$
\eqn\taude{<\p X(1) \p
X(v)>\sim-{1\over \tau^2}\sim <\psi\psi(1)\psi\psi(v)>.} With
proper normalization we get the same result as in the bosonic
case.

Thus we get the same recursion relation for ${\cal I}_n$ \recu\ as
in the bosonic string \ma; the only change is to replace
space-time dimension $26$ by $10$.

Using the expression for NS state density \eqn\sden{{\cal
G}_{NS}(N)=\oint{dw\over w}w^{-N}f_{NS} \quad ,} we get the
superstring semi-inclusive decay rate as a series summation
\eqn\sppmm{F=\sum_{p=1}^A (n-m^2(p-\half)){\cal G}_{NS}[N-np+\half
m^2(p^2-p)]+F_{NU},} with $N$ the mass level of the initial
string; $n=N-N'$, $N'$ the mass level of one of the final strings
states; $m$ the mass of the other of the final strings whose state
is fixed from the start. The above expression has the same form as
in the bosonic string.

Note that we considered only the NS sector, and in the following
we will argue in the spirit of \russ\ that NS emission is the
dominant decay channel for $N$ large, and R sector is suppressed
by $1\over N$, with $N$ being the initial string mass level. Here
for the semi-inclusive amplitude squared \amps, it is more
convenient to introduce only a projector for one of the final
states \eqn\prone{F=\sum_{ \Phi_i|_{N}}\oint {dz'\over
z'}z'^{N-N'}<\Phi_i|V(k,1)^\dagger {{1+e^{i\pi F}}\over2}
V(k,v)|\Phi_i>.} Before GSO projection, the lowest states in R
sector are already massless.
 Their vertex operators are of the form \joe\
\eqn\rvert{V^R(u^\alpha;k^\mu)=e^{-{\phi\over2}}\bar u^\alpha\Theta_\alpha e^{ikX},}
where the $\phi$ term comes from superconformal ghosts, and the $\Theta$ term, called spin fields, comes from
worldsheet fermion zero modes.
Compare this with the massless state \vortz\ in NS sector,
 we see that to get to the same mass level, NS sector needs
one more piece of excitation of the weight of $\p X$ than for R
sector. The same argument works also for the higher mass levels.

Relative to R sector, this extra excitation will contribute in \prone\ a factor of order $<\Phi_i|\p X(1)\p X(v)|\Phi_i>$
for each term in NS sector.
Ignoring dependence on the worldsheet
coordinates, $\p X\p X$ acts mainly as the energy-momentum tensor which can be seen by noting that
 the $X$ scalar contribution to the energy-momentum tensor is
\eqn\energy{T^X (z)=-{1\over\ap}\p X^\mu\p X_\mu.} So $<\Phi_i|\p X(1)\p X(v)|\Phi_i>$ is of order the
mass square of the initial string state which scales as $N$ with $N$ large. This extra factor of order $N$
makes NS sector the dominant decay channel. And our truncation to NS sector in the above is seen to
be enough at the leading order.

\newsec{Superstring decay with compactification and KKLMMT scenario}

We can also compactify type II strings on the torus. According to
the above calculations, we will obtain the same formulas as in
\ppmm, \ppm-\gl. Now the state density has the asymptotic form
\eqn\sdens{{\cal G}(N_L)\approx 2^{-{13\over4}}N_L
^{-{11\over4}}e^{\pi\sqrt{8N_L}}.} The computation of the
multiplicative constant  in the $r. h. s.$ is included in the
appendix for completeness.

The results on the toroidal compactifications can be used to
estimate the massive string decay rates in the more realistic
 KKLMMT model \kklmmt, where the type IIB strings live in a highly warped throat with Klebanov-Strassler
geometry \ks. It was shown in \recon\ how to use the string reconnection probability for toroidal
compactification to study corresponding processes in the warped geometry. We will make use of
their treatment.

A general warped geometry
\eqn\warp{ds^2=H^{-\half}(Y)\eta_{\mu\nu}dX^\mu dX^\nu +H^\half (Y)g_{ij}dY^i dY^j}
will induce a potential \recon
\eqn\poten{V(Y)={1\over{2\pi \ap H^{1\over2}(Y)}}}
on the string worldsheet, this potential confines the string to a small region
in the compact dimension near the position where this potential has a minimum. Fluctuations in these
compact dimensions \recon
\eqn\fluct{<Y^iY^i>={\ap\over2}\ln[1+{1\over{2\pi\ap^2 H^\half (0)\p_i^2 V(0)}}],
\quad \quad (\hbox{no  sum on  i} )    }
provide an effective compactification volume for strings living in. When $\p_i^2 V$ is
small in string units, that is, the geometry varies slowly on the string scale, we can look upon the
strings in such a warped geometry as in a box with the effective compactification volume, and
flat space calculations above applies.

Near the base of Klebanov-strassler solution \ks, the geometry is
locally $R^3 \times S^3$ with the warp factor
\eqn\kswarp{H(Y)=H(0)(1-{{b'r^2}\over{g_s J \ap}})} near the
origin of $R^3$. $H(0)^{-{1\over4}}$ is determined to be of order
$10^{-4}$ \recon, and $b'$ nearly 1 \hko. And $J$ counts the
number of flux.

As in \recon, write the minimum volume of a six-torus as
$V_{min}=(4\pi^2 \ap)^3$, and the effective volume of the six
transverse dimensions as $V_{\perp}$. The dimensionless quantity
$V_{min}\over V_{\perp}$ can be used to measure the effects of
compactification. Quantum fluctuations on the $R^3$ contribute
${(2\pi)^{3\over2}}\over{\ln^{3\over2}(1+g_sJ)}$ to $V_{min}\over
V_{\perp}$ \recon.
 For $S^3$, quantum fluctuations give ${(2\pi)^{3\over2}}\over{\ln^{3\over2}(H^{\half}(0))}$,
 and its volume constrains the factor to be no less than
 ${4\pi}\over{(g_sJ)^{3\over2}}$ \recon. So finally we get
\eqn\vvratio{{V_{min}\over V_{\perp}}\approx \inf
\big[{{{4\pi}\over{(g_sJ)^{3\over2}}}{{(2\pi)^{3\over2}}\over{\ln^{3\over2}(1+g_sJ)}}}
,
{{(2\pi)^3}\over{\ln^{3\over2}(H^{\half}(0))\ln^{3\over2}(1+g_sJ)}}
\big ].}

Here we need not to worry about winding modes, and as a rough estimate we can also
neglect the contribution from KK modes. The flat metric total decay rate \ttrate\ for
 a typical string of mass $M$ to
two strings with mass $m$ and $M_2$, when these two modes are
excluded, can be simplified \eqn\kkrate{\Gamma [M\to m+M_2]
\approx A_{D-d_c}{1\over V_{\perp}}{g_c^2\over M^2}{\cal N}^2
{\cal G}^2 k^{D-3-d_c},} with ${\cal N}={\cal N}_L={\cal N}_R$,
and ${\cal G}={\cal G}_L={\cal G}_R$, and compactification volume
$V_{\perp}=\prod _i ^{d_{c}} R_{i}$. Remember that
 the initial string is averaged, and
the final two strings both sum over given mass level. For the
warped geometry, \kkrate\ represents results seen by a
10-dimensional observer. And the quantities measured by the
4-dimensional observer, in our case the mass and the momentum, get
red-shifted. Thus we should multiply $M, m,M_2$ and $k$ all by a
red-shift factor $H^{1\over4}(0)$. Now the mass shell conditions
read like \eqn\ksmash{\ap M^2H^{\half}(0)=4(N-1).}

Let's look at the eq. \gggam. Now we need to replace $t$ by $H^{1\over4}(0)t$
and $m_R$ by $H^{1\over4}(0)m_R$ to get $\Gamma^k$ replaced by $\Gamma^k H^{-{15\over8}}(0)$, with $D=10$ and
$d_c=6$ explicitly inserted.

Then the emission rate of a string with mass $m$ from a typical string of mass $M$ in the Klebanov-Strassler
geometry can finally be estimated to be
\eqn\kkresult{{{d\Gamma_{KS}(m)}\over{dm}}\approx A_{KS}g_c^2 H^{-{13\over8}}(0)Mm_R^{-{15\over2}},}
with $A_{KS}= {1\over{16\pi^5}}{{V_{min}}\over{V_{\perp}}}$.

With $H^{-{1\over4}}(0)\sim 10^{-4}$, we see that the emission
rate is suppressed by a factor of order $10^{-26}$ in this warped
geometry! With $A_{KS}$ no larger than $10^{-4}$, $g_S$ no larger
than unity, the multiplicative constant in \kkresult\ is no larger
than $10^{-30}$. Thus in this warped geometry, the decay channels
for a massive string to two massive strings are greatly
suppressed.

We note that the situation is different if one of the two final strings is massless. It is found \russ\ that the decay rate is
proportional to the mass of the initial string. So including the redshift factor, it can be written as
\eqn\massless{\Gamma_{massless}\sim g_c^2 M H^{1\over 4}(0),}
from which we see that this decay channel is enhanced by order $10^4$. Thus in such warped geometries,
massive strings may lose their energy mainly through gravitational radiation.

\newsec{Discussions}

We calculated the average decay rate of a typical massive string state in
superstring theory, generalizing the earlier results obtained in the bosonic
string theory. We did this calculation with a torus compactification, which
at the first look is not realistic enough. However, we see that the calculation
can be applied to the brane-inflation model of KKLMMT, we find that the decay
is greatly suppressed. We believe that this shall have important phenomenological
implications.

To apply our calculation to the problems concerning cosmic strings, we need
to know how fundamental strings are generated, for instance in the process
of brane-anti-brane annihilation. With that knowledge, we can write down a set
of differential equations on the distribution function of massive strings, making
use of our result. We may need to take the evolution of the universe into account
too. The approach outlined here may replace the usual simulation process to
predict the number and the distribution of massive cosmic strings that remain
to this day in the sky.

Particularly, it has been argued in \gwvil\ that gravitational
radiation is not an efficient enough energy loss mechanism for the
string network to be consistent with present observation. Other
mechanisms are needed to prevent the network from quickly dominate
the universe. And from our calculations above, we can see that the
large suppression factor is a generic feature for string
interactions in the warped geometry with three massive strings
involved. Thus it is important to study more carefully the energy
loss mechanisms for F-D string networks in warped geometries.

{\bf Acknowledgments}

This research project was supported by a grant from CNSF and a
grant from CAS. We are also grateful to the Interdisciplinary
Center of Theoretical Studies for constant support. BC would like
to thank C.J. Zhu for the clarification on the chiral splitting of
the massive superstring state.

\appendix{A}{ Compute the state density for superstrings.}
The asymptotic densities of the states for bosonic and superstrings are standard
textbook knowledge (see for example \gsw\
and \joe). However, in our computation we need to know the detailed numerical coefficients for superstrings
(Manes has done for bosonic string in \ma),
 which are often not explicitly demonstrated, so we include this appendix to write them out. We follow the whole procedure
 of \gsw.

Consider first open strings and NS sector only. The state
degeneracy ${\cal G}_{NS}(n)$ is given by
\eqn\nsdeg{f_{NS}(w)={\rm Tr}{{1+e^{i\pi
F}}\over2}w^N=\sum_{n=0}^{\infty}{\cal G}_{NS}(n)w^n8\prod_{n=1}^{\infty}({{1+w^n}\over{1-w^n}})^8,} with $N$ the
summation of the bosonic and fermionic number operators.
Generalization of Hardy-Ramanujan formula gives
\eqn\hr{\prod_{n=1}^{\infty}({{1+w^n}\over{1-w^n}})^{-1}=\vartheta_4
(0|w)=(-{\ln w\over\pi})^{-{\half}} \vartheta_2(0|e^{\pi\over \ln
w}),} where the modular transformation of $\vartheta$ function
\eqn\thethe{\vartheta_4 (0|\tau)=(-i\tau)^{-\half}\vartheta_2
(0|-{1\over\tau})} has been used, with \eqn\tttau{\tau=-{{i\ln
w}\over{\pi}}.}

As $w\to1$, the second argument of $\vartheta_2$, which now reads \eqn\taupr{\tau'=-{1\over\tau}=-{i\pi\over{\ln w}},}
approaches $\infty$.
We know from the expansion \eqn\extht{\vartheta_2 (0|\tau')=\sum_{n=-\infty}^{\infty}e^{i\pi (n-\half)^2\tau'}} that
\eqn\thetwo{\vartheta_2(0|\tau'\to\infty)\to2e^{{i\pi\over4}\tau'}=2e^{{\pi^2\over{4\ln w}}}.}
Thus \hr\ is asymptotically
\eqn\ffns{\prod_{n=1}^{\infty}({{1+w^n}\over{1-w^n}})^{-1}\to(-{\ln w\over\pi})^{-{\half}}2\exp({{\pi^2\over{4\ln w}}}).}

>From \nsdeg, the state degeneracy ${\cal G}_{NS}(n)$ can be expressed as a
contour integral on a small circle around $w=0$
\eqn\degns{{\cal G}_{NS}(n)={1\over 2\pi i}\oint{f_{NS}\over w^{n+1}}dw.}
To compute the above integration, we make a saddle point approximation near $w=1$. The power of $w$
can be put on the exponential
\eqn\sad{{\cal G}_{NS}(n)={1\over 2\pi i}\oint 8 (-{\ln w\over\pi})^4 2^{-8}\exp[{-{{2\pi^2}\over{\ln w}}
-(n+1)\ln w}]dw,}
to get the saddle point at
\eqn\sadpt{\ln w_0={{\sqrt2\pi}\over\sqrt{n+1}},}
where expansion can be made
\eqn\sadexp{\ln w=\ln w_0+iu.}
Then ${\cal G}_{NS}(n)$ is approximately
\eqn\sadappx{{\cal G}_{NS}(n)\sim {1\over{2\pi }}{1\over32}({\sqrt2\over\sqrt n})^4 e^{\pi\sqrt{8n}}
\int_{-\infty}^{\infty} \exp(-{{\sqrt2 n^{3\over2}}\over\pi}u^2)du.}
Carrying out the integration over $u$ we find
\eqn\saddle{{\cal G}_{NS}(n)\sim2^{-{13\over4}}n^{-{11\over4}}e^{\pi\sqrt{8n}}.}
Or using $n\sim \ap m^2$, write it out in terms of mass
\eqn\degmass{{\cal G}_{NS}(m)\sim 2^{-{13\over4}}{\ap}^{-{11\over4}}m^{-{11\over2}}e^{\pi\sqrt{8\ap}m}.}
 Here we use the convention ${\cal G}_{NS}(m)dn={\cal G}_{NS}(n)dn$, different from \gsw.

At this point, we also note that R sector has the same expression. And combine the left
and right pieces together we arrive at the expression
for closed strings
\eqn\degcl{{\cal G}^{cl}(n)=[{\cal G}^{op}(n)]^2\sim 2^{-{9\over2}}n^{-{11\over2}}e^{4\pi\sqrt{2n}}.}
Taking care of the difference between the mass shell conditions of open and closed strings ($\ap m^2\sim 4n$
for closed strings),
the state degeneracy for closed string as a function of mass reads
\eqn\degmcl{{\cal G}^{cl}(m)\sim 2^{13\over2}{\ap}^{-{11\over2}}m^{-11}e^{\pi\sqrt{8\ap}m}.}
Thus open and closed strings have the same Hagedorn temperature
\eqn\hag{T_H={1\over{\pi\sqrt{8\ap}}}.}

\listrefs
\end